\documentclass[pre,twocolumn,showpacs,floatfix,superscriptaddress,aps]{revtex4}
\usepackage{amsmath}
\usepackage{makeidx}
\usepackage{amssymb}
\usepackage{graphicx}

\usepackage[usenames]{color}
\usepackage[normalem]{ulem}

\newcommand{\beq}{\begin{equation}}
\newcommand{\eeq}{\end{equation}} 
\newcommand{\beqa}{\begin{eqnarray}}
\newcommand{\eeqa}{\end{eqnarray}} 
\begin{document}


\title{Stable spatial and spatiotemporal optical soliton  in the   core of an optical vortex}

\author{ S. K. Adhikari\footnote{adhikari@ift.unesp.br; URL: http://www.ift.unesp.br/users/adhikari}
} 
\affiliation{
Instituto de F\'{\i}sica Te\'orica, UNESP - Universidade Estadual Paulista, 01.140-070 S\~ao Paulo, S\~ao Paulo, Brazil
} 

\begin{abstract}

We demonstrate a robust, stable, mobile, two-dimensional (2D)
spatial and  three-dimensional (3D) spatiotemporal optical soliton 
in  the core of an optical vortex, while all nonlinearities are of the 
 cubic (Kerr) type.  
The 3D  soliton  can  propagate with a constant velocity along the vortex core   
without any  deformation.  Stability of the soliton under a small perturbation is established numerically. 
Two such solitons  moving along the vortex core can undergo a quasi-elastic collision at medium velocities.  Possibilities of forming such a 2D
spatial soliton in the core of a vortical beam are discussed.

\end{abstract}


\maketitle

 \section{Introduction}
 
A bright soliton is a self-bound object that travels at a constant velocity  
in one dimension (1D), due to a cancellation of  nonlinear attraction and defocusing forces \cite{book,rmp,sol}. 
An 1D dark soliton is a dip in  uniform density, which also moves with a constant velocity maintaining its shape \cite{book,dark}. The 1D soliton  has been observed  in nonlinear optics \cite{book,sol} and in Bose-Einstein condensate (BEC) \cite{rmp}. 
 Specifically, optical temporal solitons were observed by Di Trapani {\it et al.}
\cite{temporal} for a cubic Kerr nonlinearity.
However, a three-dimensional (3D)   spatiotemporal soliton
cannot be formed in isolation with a cubic Kerr nonlinearity 
due to collapse \cite{collapse,book}. 
The same is true about  a two-dimensional (2D)  spatial soliton  with a Kerr nonlinearity.
Under special condition a 2D   spatiotemporal optical soliton has 
been observed \cite{stempo}. 
However, the solitons  can be stabilized in higher dimensions
 for  a saturable or a modified nonlinearity 
\cite{arrest,modnon}, or by a nonlinearity \cite{sadhan,malomed1} or dispersion \cite{sadhan2} management  among other possibilities \cite{malomed}.

In this paper we 
demonstrate the formation of a 2D  spatial and  a  3D spatiotemporal
bright  soliton  with Kerr nonlinearity in the   core of an optical beam hosting a quantized vortex (vortical beam) in 
a   Kerr medium, which we  call  
 a binary optical vortex-soliton. A 2D spatial vortex in an infinite  repulsive Kerr medium has been experimentally observed \cite{vorop} and theoretically studied 
\cite{vorth}. A 3D spatiotemporal optical vortex in an infinite  repulsive Kerr medium suffers from transverse instability and vortex-line bending
similar to a 3D superfluid vortex \cite{pita}. However, in a binary 
beam
the vortex-soliton is found to be dynamically stable under a small perturbation not only in the   2D spatial case, but also in the  3D spatiotemporal case. 
In our simulation no transverse instability of the vortex line is noted.
The nonlinear interaction between the optical soliton and the vortical beam is always taken to be repulsive. This mutual repulsion stabilizes the soliton in the $x-y$ plane and also prevents the vortex line from bending.  
The soliton  can swim freely along the $t$ direction  with a constant velocity  along the core of the vortical beam.
  Because of a strong localization 
of the soliton  due to inter-beam repulsion, the soliton  can move without visible deformation along the vortex core. 
  At medium velocities the collision between two such solitons   
is found to be quasi-elastic.   

A related study by Law {\it et al.} \cite{law} of stable vortex-soliton structure in two-component BEC bears some formal similarity with the present study.
However, that  study should be considered to be complimentary to the present study 
rather than overlapping. In the present study in nonlinear optics there are no traps, whereas in Ref.
\cite{law} trapped BEC has also been considered.  
In the 2D case, they \cite{law} consider only repulsive nonlinearity in the soliton whereas we consider both focusing (attractive) and defocusing (repulsive) nonlinearities in the soliton and establish stable solitons in both cases.  The 2D vortex-soliton with focusing nonlinearity was predicted in Ref. 
\cite{mus} from an analytic consideration. However, no stability analysis of these 
vortex-solitons was presented in Ref. \cite{mus} for Kerr nonlinearity, whereas we present convincing numerical tests of stability.
In the 3D case,  Law {\it et al.} \cite{law} present numerical results for the trapped case only, whereas   the present results are obtained in the absence of any trapping potential. In the 3D trapless case, they argue 
in favor of  stable     vortex-soliton structure with repulsive nonlinearity in the soliton, whereas we find stable   vortex-soliton structure only for attractive nonlinearity in the soliton. 
Further studies of the 2D vortex-bright-soliton structure of Ref. \cite{law} are presented in Ref. \cite{pola}.

 The  vortex-soliton in higher dimensions is a generalization of the 1D optical dark-bright soliton \cite{dark-bright,book}. 
 Hence
we present the nonlinear Schr\"odinger  (NLS) equations used in this study in 2D and 3D in Sec. 
\ref{II} together with a discussion of the 1D optical dark-bright soliton.
The numerical procedure for including a vortex in a uniform system is explicitly 
presented.  
In Sec. \ref{III} we present the numerical results for stationary profiles of 2D spatial and 3D spatiotemporal vortex-solitons.  We present numerical tests of stability of the vortex-soliton under a small perturbation. 
The quasi-elastic nature of collision 
of two solitons moving along the vortex core is also established. 
We end with a summary of our findings in Sec. \ref{IV}.

\section{Nonlinear Schr\"odinger equations}
 
\label{II}

    A vortex in an   optical beam \cite{vorth,vorop}  bears similarity with 
a 1D dark soliton in generating a hole along the axial $t$ direction and is often called a 3D dark soliton \cite{book}. 
Hence, the present  binary optical vortex-soliton is the 3D analogue of the well-known  1D dark-bright soliton \cite{dark-bright}. 
To understand how  a  3D vortex-soliton can appear,  we consider the following integrable binary 1D
dark-bright soliton  model in all-repulsive Kerr medium, in the form of a coupled 
NLS equation 
\begin{equation}\label{model}
 {\Big [} i \frac{\partial }{\partial z}  +\frac{1}{2 }\frac{\partial^2}{\partial t^2}
-\sum_i\vert \psi_i \vert^2 {\Big ]}  \psi_j(t,z)=0, 
\end{equation}
in  scaled  units   where $i,j=1,2,$ represents the dark and the bright solitons, respectively.   The solitons of Eq. (\ref{model}) are 1D temporal 
while $t$ and $z$ denote time and space variables, respectively. When the time variable $t$ is replaced 
by the spatial variable $x$ the model becomes 1D spatial.
Equation (\ref{model})
hosts the analytic dark-bright soliton \cite{book} 
\begin{align} \label{x1}
\psi_1(t,z)= & \beta \tanh[\alpha(t-vz)]e^{ivt-i(v^2/2+\beta^2) z},\\
\psi_2(t,z)=&\gamma\mathrm{sech}[\alpha(t-vz)]  e^{ivt+i[(\alpha^2-v^2)/2-\beta^2]z},
\label{x2}
\end{align}
where $\alpha$ and $\beta$ $(\beta>\alpha)$ are  constants which control the intensity and width of the solitons,  $\gamma=\sqrt{\beta^2-\alpha^2} $ and 
$v$ determines the  velocity.   The  bright soliton (\ref{x2}) is formed in the 
all-repulsive Kerr model  (\ref{model}) 
 due to the accompanying  dark soliton (\ref{x1}).

The 1D bright soliton (\ref{x2}) stays in the central hollow of the dark soliton
(\ref{x1})  and is confined due to the repulsive nonlinearity 
between the (outer) dark   and (inner) bright solitons.     Similarly, the soliton  of a 2D spatial  or a 3D spatiotemporal 
binary  vortex-soliton 
can be  confined in the radial $x-y$ plane by the  repulsion between the vortex and   the soliton. In the 2D spatial case, the confinement of the soliton in the vortex-soliton is possible for a moderately self-focusing nonlinearity or all self-defocusing nonlinearity 
in the soliton. For   a large  self-focusing Kerr nonlinearity the soliton
collapses \cite{arrest}. In the 3D spatiotemporal case, a stable soliton in the vortex-soliton 
can only be obtained   provided  we consider a weak self-focusing Kerr nonlinearity  in the bright soliton. The soliton escapes to infinity for a self-defocusing  
   Kerr nonlinearity in the soliton and collapses for a large self-focusing Kerr nonlinearity \cite{modnon}.





For the formation of a  3D  spatiotemporal vortex-soliton we consider  
the following binary dimensionless NLS equations  with self-focusing nonlinearity in the soliton
    \cite{book}
\begin{align}& \,
{\Big [}  i \frac{\partial }{\partial z} +\frac{\nabla_\perp^2}{2 }+\frac{1}{2}\frac{\partial^2}{\partial t^2}-  \vert \phi_1 \vert^2
-  \vert \phi_2 \vert^2
{\Big ]}  \phi_1({\bf r},z)=0,
\label{eq4}
\\
& \,
{\Big [}  { i} \frac{\partial }{\partial z}
+  \frac{\nabla_\perp ^2}{2}+\frac{1}{2}\frac{\partial^2}{\partial t^2}
+ \vert \phi_2 \vert^2 
-  \vert \phi_1 \vert^2  
{\Big ]}  \phi_2({\bf r},z)=0,
\label{eq5}\\
& \nabla_\perp ^2= \frac{\partial^2}{\partial x^2}+
\frac{\partial^2}{\partial y^2},
\end{align}
in scaled units 
where ${{\bf r} \equiv \{x,y,t\}}$. Both in 2D and 3D the first component $i=1$ will host the vortex and the second component  $i=2$ will host the soliton.
The numerical simulation is performed in a cubic box (of length $2L$) limited by $|x|,|y|,|t| <L$.
The beams have powers $P_i$ defined by 
$P_i=\int_{|x|,|y|,|t| <L} d{\bf r} |\phi_i({\bf r})|^2$. In the limit $L\to \infty$ the power $P_1$ diverges. This is not of concern. The vortex-soliton is  controlled by the finite power density $p_1\equiv P_1/(2L)^{\cal D}$ of the vortex, where  ${\cal D}$ is the dimension: ${\cal D}=2$ for 2D and = 3 for 3D. We will classify the vortex states by their finite power density $p_1$.
There are three nonlinearities in this 
binary optical system, two of which are fixed by the powers $P_i$, and the third by 
the length scale, thus making Eqs. (\ref{eq4}) and (\ref{eq5}) free of parameters.  
The plus sign before $|\phi_2|^2$ in Eq. (\ref{eq5})
denotes a self-focusing nonlinearity in component 
2 which will host the soliton. All other nonlinearities with a negative sign 
denote self-defocusing.
 
Similarly, a 2D spatial  vortex-soliton is described by the following scaled NLS equations 
\begin{align}& \,
{\Big [} i \frac{\partial }{\partial z}  +\frac{\nabla_\perp^2}{2 }-  \vert \phi_1 \vert^2
-  \vert \phi_2 \vert^2
{\Big ]}  \phi_1({\bf r},z)=0,
\label{eq6}
\\
& \,
{\Big [}  { i} \frac{\partial }{\partial z}
+  \frac{\nabla_\perp ^2}{2}
\pm \vert \phi_2 \vert^2 
-  \vert \phi_1 \vert^2  
{\Big ]}  \phi_2({\bf r},z)=0,
\label{eq7}
\end{align}
  obtained after removing the time variable from Eqs.(\ref{eq4})-(\ref{eq5}),  where now ${{\bf r} \equiv \{x,y\}}$. The numerical simulation in this case is performed in a square  (of length $2L$). The powers  $P_i$ in this case are defined by 
$P_i=\int_{|x|,|y| <L} d{\bf r} |\phi_i({\bf r})|^2$.
The plus  sign in Eq. (\ref{eq7}) before $ \vert \phi_2 \vert^2$
correspond to a self-focusing nonlinearity and the  minus  sign to a self-defocusing nonlinearity.

  To find a stationary quantized vortex of charge   ${\cal C}$ 
in component 1,  also called a dark
soliton with circular symmetry,
 we look for circularly-symmetric spatiotemporal solution $\Phi_1 ({\bf r},z)$
in $x-y$ plane:  
$\phi_1({\bf r},z) \equiv \Phi_1 ({\bf r},z)e^ {i{\cal C}\varphi} $,
where $\varphi$ is the azimuthal angle  and  $\Phi_1({\bf r},z)$ satisfies \cite{vorop,vorth}
\begin{align}& \, 
{\Big [}  i \frac{\partial }{\partial z} +\frac{1}{2} \frac{\partial^2}{\partial t^2 }+\frac{\nabla_\perp^2}{2} -\frac{{\cal C}^2}{2(x^2+y^2)} - \vert \Phi_1 \vert^2
-  \vert \phi_2 \vert^2
{\Big ]}\Phi_1 =0,
\label{eq9}
\end{align}
 with  boundary conditions
$\Phi_1 (x=0,y=0,t) = 0,
\Phi_1 (x\to \infty,y\to \infty, t) =$ constant \cite{vorth}. 
 The boundary conditions  on the 1D dark soliton 
(\ref{x1}) are very similar: $\psi_1(t=0)=0, \psi_1(t\to \infty)=$ constant.
In the 2D spatial case the time derivative in Eq. (\ref{eq9})  is dropped and one has 
\begin{align}& \, 
{\Big [}  i \frac{\partial }{\partial z} +\frac{\nabla_\perp^2}{2} -\frac{{\cal C}^2}{2(x^2+y^2)} - \vert \Phi_1 \vert^2
-  \vert \phi_2 \vert^2
{\Big ]}\Phi_1 =0,
\label{eq10}
\end{align}
 with  boundary conditions
$\Phi_1 (x=0,y=0) = 0,
\Phi_1 (x\to \infty,y\to \infty) =$ constant \cite{vorth}.
For a bright soliton  of component 2 in the vortex core of component 1 we solve Eqs. (\ref{eq9}) and (\ref{eq5}) in the 3D spatiotemporal case 
or Eqs. (\ref{eq10}) and (\ref{eq7}) in the 2D spatial case.  We take ${\cal C}=1$ in this paper. Vortices with charge ${\cal C} >1$ are usually unstable and decay into two vortices of unit charge.

\section{Numerical Results}

\label{III}

Unlike in the 1D case, the coupled NLS equations for the 
2D  spatial and 3D     spatiotemporal binary vortex-soliton do not have analytic solution and we solve them numerically
by the split-step 
Crank-Nicolson method using both real- and imaginary-$z$ propagation
  in Cartesian coordinates  
using a ${\bf r}$  step of  $ 0.2$
and a $z$ step of  $ 0.0025$ \cite{CPC}.

\begin{figure}[!t]

\begin{center}
\includegraphics[width=\linewidth,clip]{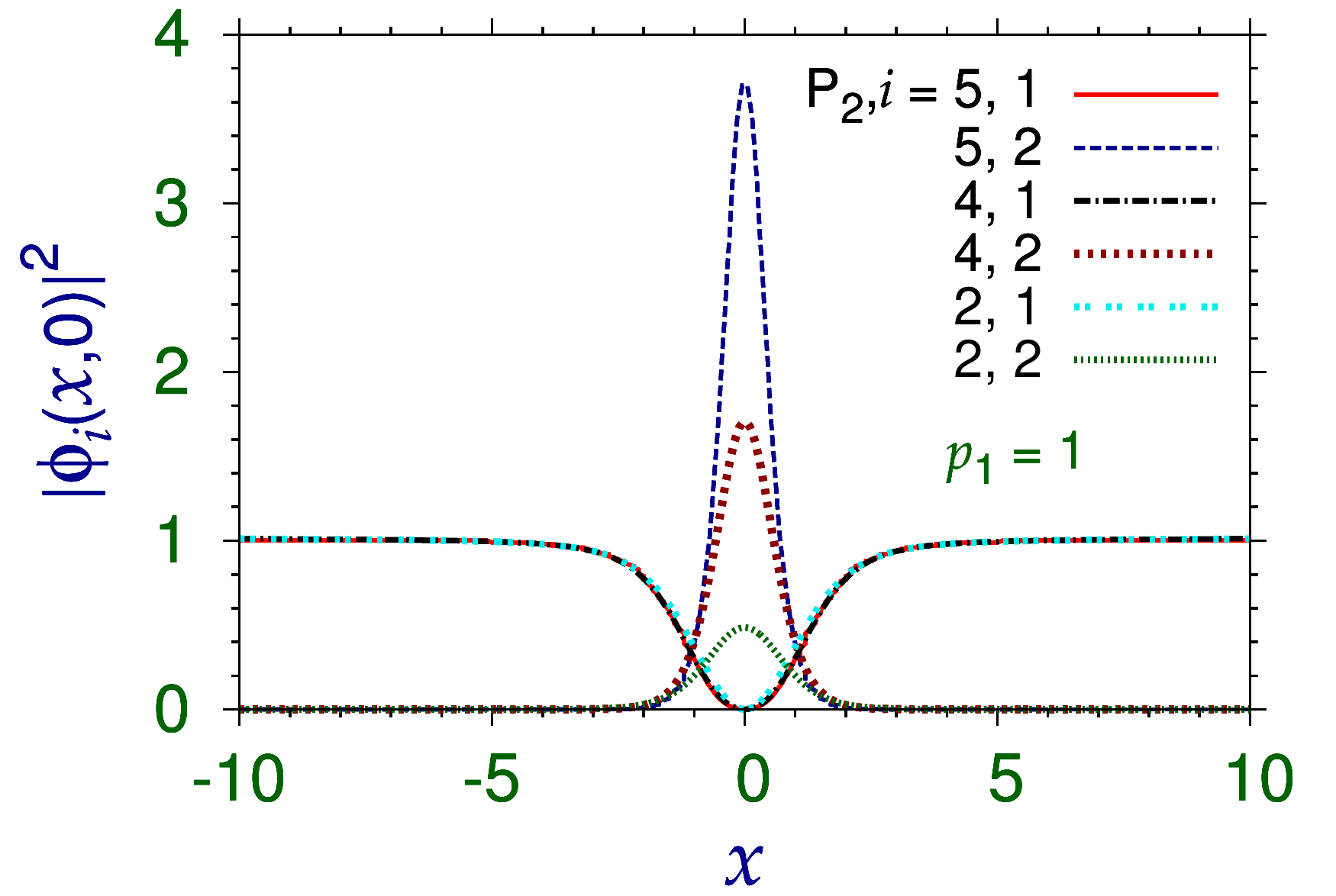} 

\caption{ (Color online) Vortex  ($i=1$) and soliton ($i=2$) densities $|\phi_i(x,0)|^2$ of the 
2D  spatial vortex-soliton with vortex power density  $p_1=1$ and soliton power $P_2=2,4,5$, from a solution of Eqs. (\ref{eq7}) and (\ref{eq10})
for  a self-focusing nonlinearity [$+$ sign before $|\phi_2|^2$ in Eq. (\ref{eq7})]  in the soliton.
}\label{fig1} \end{center}

\end{figure}

We solve  Eqs. (\ref{eq7}) and (\ref{eq10})  for the  2D spatial case and Eqs. (\ref{eq5}) and (\ref{eq9}) for the 3D spatiotemporal case. In both cases the charge 
in Eq. (\ref{eq9}) is unity: ${\cal C}=1$.  
In the imaginary-$z$ propagation the initial vortex state was taken as $\Phi_1({\bf r})
\sim [1-\exp\{-\alpha(x^2+y^2)  \}]$ with power $P_1$. The initial  2D
spatial soliton was taken as 
$ \phi_2({\bf r})\sim  \mathrm{sech}
(x\beta)\mathrm{sech}(y\beta)$  and the 3D spatiotemporal
soliton as $ \phi_2({\bf r})\sim  \mathrm{sech}
(x\beta)\mathrm{sech}(y\beta)\mathrm{sech}(t\gamma)$
with power   $P_2$,  where $\alpha,\beta$ and $\gamma$ are    parameters.
For a quick convergence   these  parameters   should be chosen conveniently   so that these states are good approximations to the final states.  The change of power from one simulation to another 
is obtained by varying the amplitude of the input beam as the width of the same is held 
constant throughout the study.
The powers $P_i$ are conserved quantities during numerical simulation and have the 
same value for all $z$ in 2D spatial and 3D spatiotemporal cases.
  These initial states approximate  well the vortex core and the soliton.

\begin{figure}[!t]

\begin{center}
\includegraphics[width=\linewidth,clip]{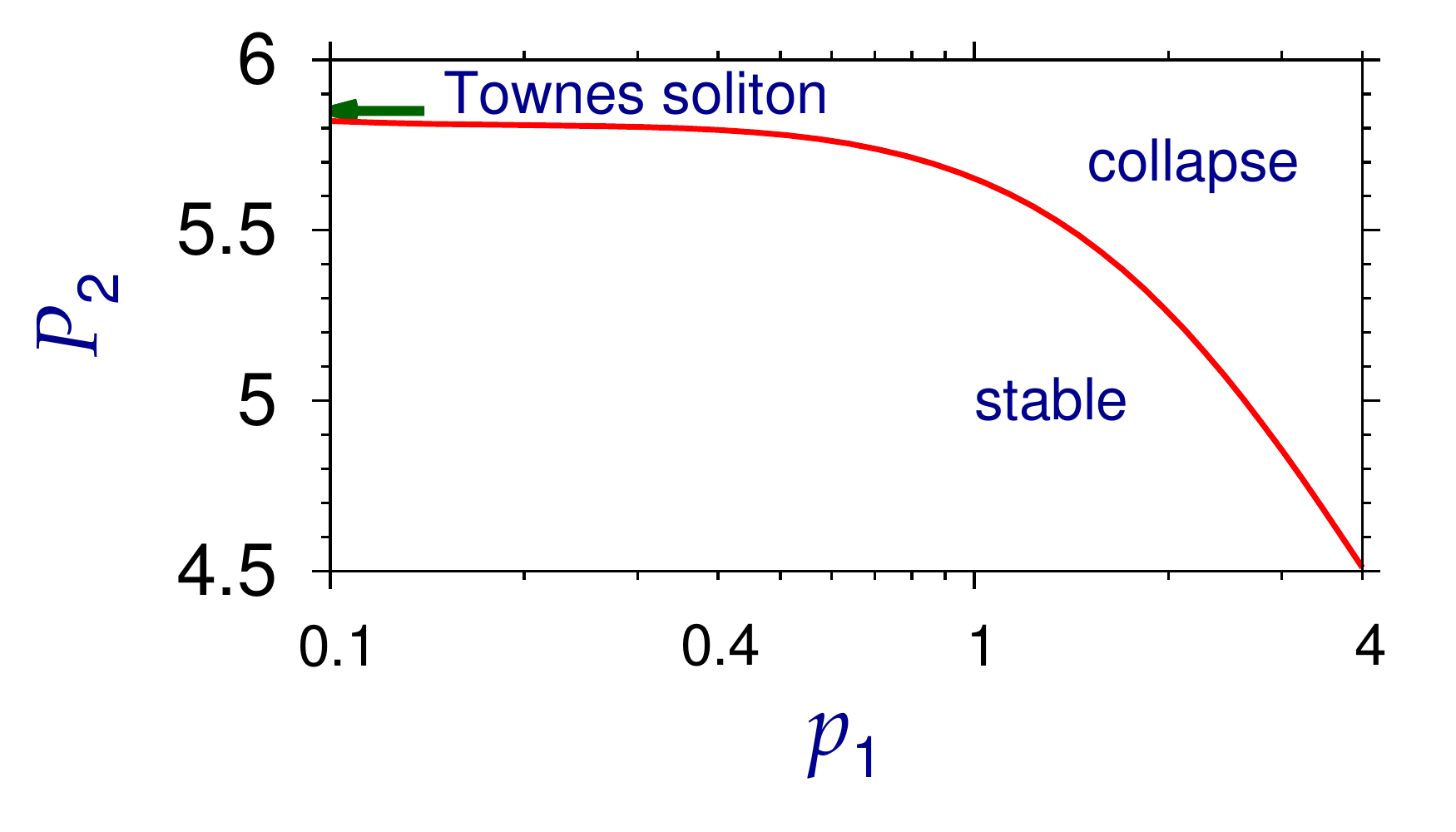} 

\caption{ (Color online) Critical value of $P_2$ for a stable 2D spatial 
vortex-soliton of Fig. \ref{fig1} with   a self-focusing nonlinearity in the soliton
obtained  from a solution of Eqs. (\ref{eq7}) and (\ref{eq10})
for 
different power density $p_1$ of the vortex. 
}\label{fig2} \end{center}

\end{figure}

First we report  results for a self-focusing nonlinearity in the soliton for a 2D spatial vortex-soliton.   The numerical simulation for the 2D spatial vortex-soliton  is performed in the square limited by  $x=y \equiv \pm L, L=\sqrt{250}$ with 
powers $P_1=1000$ and $P_2=2,4, 5$ for a self-focusing nonlinearity in the soliton and with powers $P_1=1000$ and $P_2=1, 100,100$ 
and with powers $P_1=100$ and $P_2=10, 100,1000$ 
for a self-defocusing nonlinearity in the soliton, such that ${\int_{-L}^{L} dx \int_{-L}^{L}dy} |\phi_i(x,y)|^2 =P_i$.  
The size of the vortex core is much smaller than the extension of the beam in the $x-y$ plane. 
For the  power of the vortex (in the first component) at $P_1=1000$,   the  power density $p_1\equiv P_1/(2L)^2=1$.  
In the imaginary-$z$ routine the normalization of the functions $\phi_i$ are reset to the predetermined powers $P_i$ after every $z$ iteration. For real-$z$ routine these normalizations are conserved after every $z$ iteration. 
The densities in this case, obtained by imaginary-$z$ propagation, have a circularly symmetric profile. Hence,  we present the result for density $|\phi_i(x,0)|^2$ 
only along the $x$ axis   for soliton powers $P_2=2,4,5$ in Fig. \ref{fig1}. 
For a small value 
of self-attraction in the soliton, corresponding to a power of $P_2=2$, the soliton has a weak localization corresponding to a small peak in density as can be seen in Fig.
\ref{fig1}.  For a larger power $P_2=5$,  the soliton has a  stronger localization
corresponding to a high peak in density as can be seen in Fig.
\ref{fig1}. The soliton collapses if the self-attraction in the soliton is further 
increased to $P_2=6$. The vortex profile remains practically unchanged in this case 
for different powers $P_2$ of the soliton. We also studied the stability of the soliton 
for different values of power density $p_1$ and power $P_2$ and illustrate the result  in Fig. \ref{fig2}, where we plot the critical power $P_2$ for obtaining a stable soliton for different $p_1$.  {In the absence of the vortex ($p_1=0$), in 2D, the self-focusing system (component 2) can have an unstable Townes soliton \cite{townes} of power 
$P_2=5.85$ \cite{malomed1,berge}. For powers $P_2>5.85$, the Townes soliton has an excess 
of attraction and it collapses, whereas for powers $P_2<5.85$, the attraction is too weak to bind the soliton and it escapes to infinity. A vortex with a small non-zero power density 
$p_1\to 0$ has no effect on the collapse of the Townes soliton for  power 
$P_2>5.85$, but  it arrests its uncontrolled expansion for power $P_2<5.85$ and forms a stable soliton, viz. Fig. \ref{fig2}. 
For larger power density $p_1$ of the vortex, the binding force on the soliton increases and it becomes more vulnurable to collapse 
and hence the region of collapse increases in the $P_2-p_1$ phase plot  of Fig. \ref{fig2}.}

\begin{figure}[!t]

\begin{center}
\includegraphics[width=\linewidth,clip]{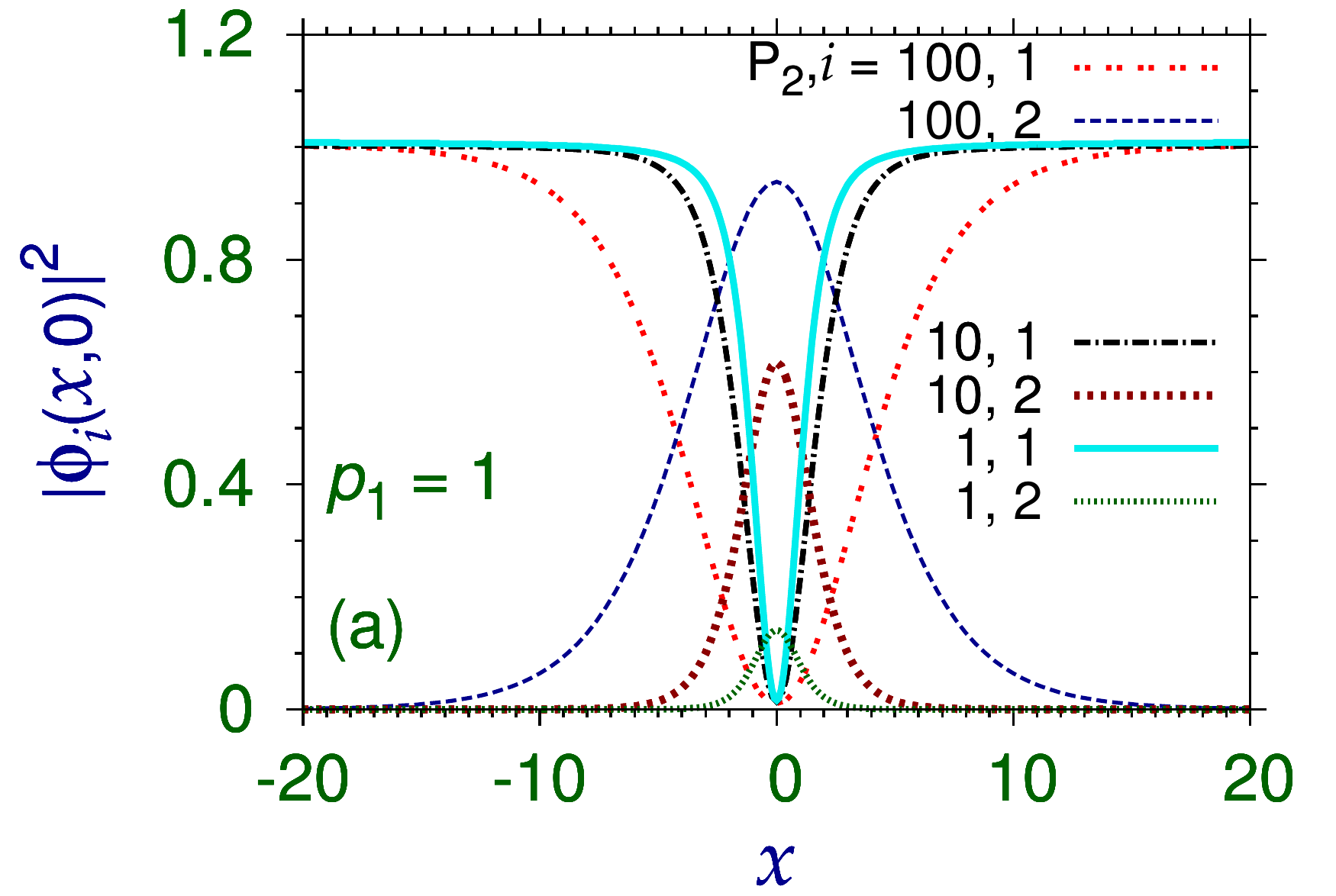}
 \includegraphics[width=\linewidth,clip]{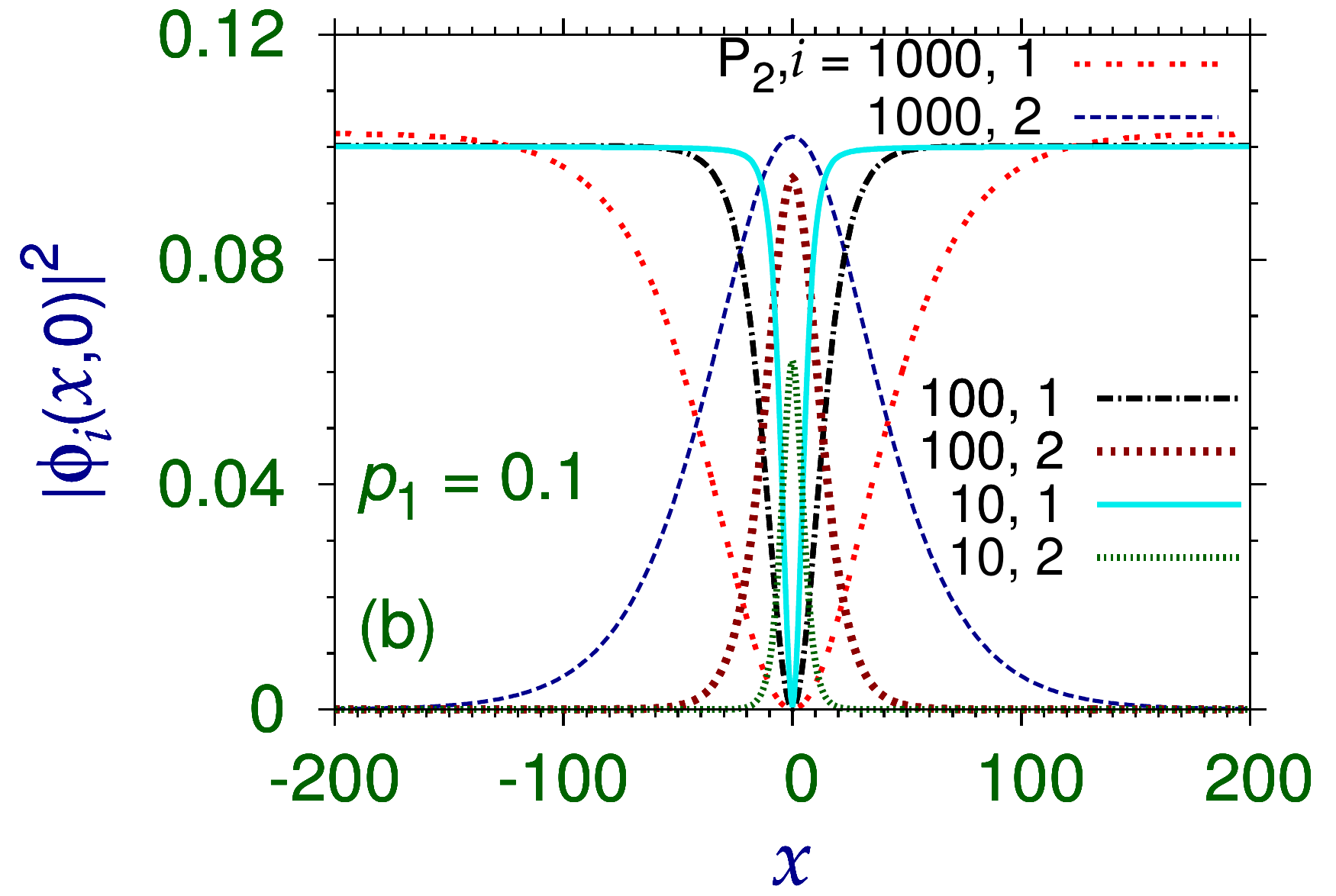}
 
\caption{(Color online) Vortex  ($i=1$) and soliton ($i=2$) densities $|\phi_i(x,0)|^2$ of the 
2D  spatial vortex-soliton with {(a)  vortex
power density $p_1 = 1$ and soliton power 
 $P_2=1,10,100$ and with (b) vortex
power density $p_1 = 0.1$ and soliton power 
 $P_2=10, 100, 1000$,} from a solution of Eqs. (\ref{eq7}) and (\ref{eq10}) 
for  a self-defocusing nonlinearity [$-$ sign before $|\phi_2|^2$ in Eq. (\ref{eq7})]  in the soliton.
}\label{fig3} 

\end{center}

\end{figure}

\begin{figure}[!t]

\begin{center}
\includegraphics[width=.8\linewidth,clip]{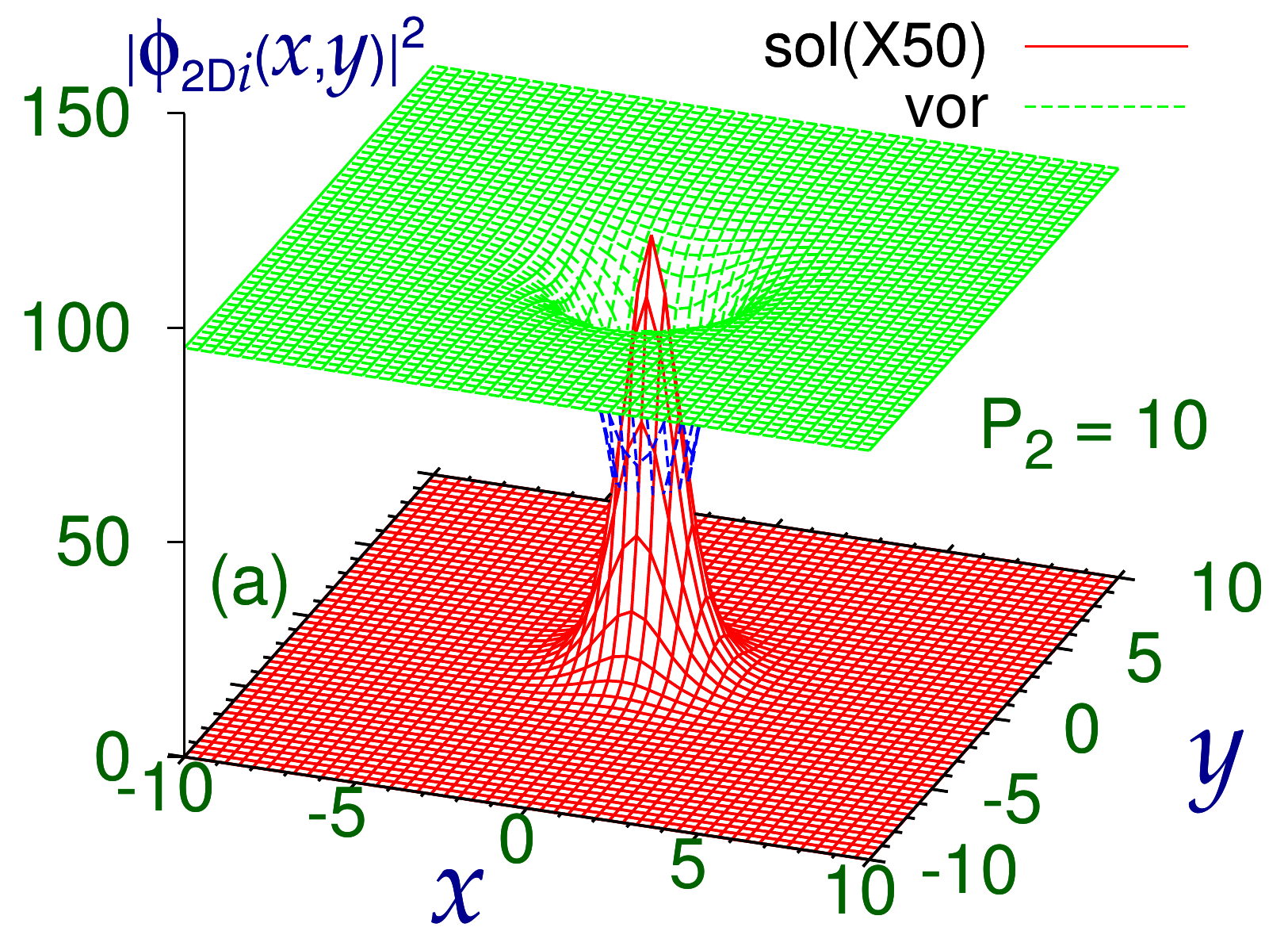}
  \includegraphics[width=.8\linewidth,clip]{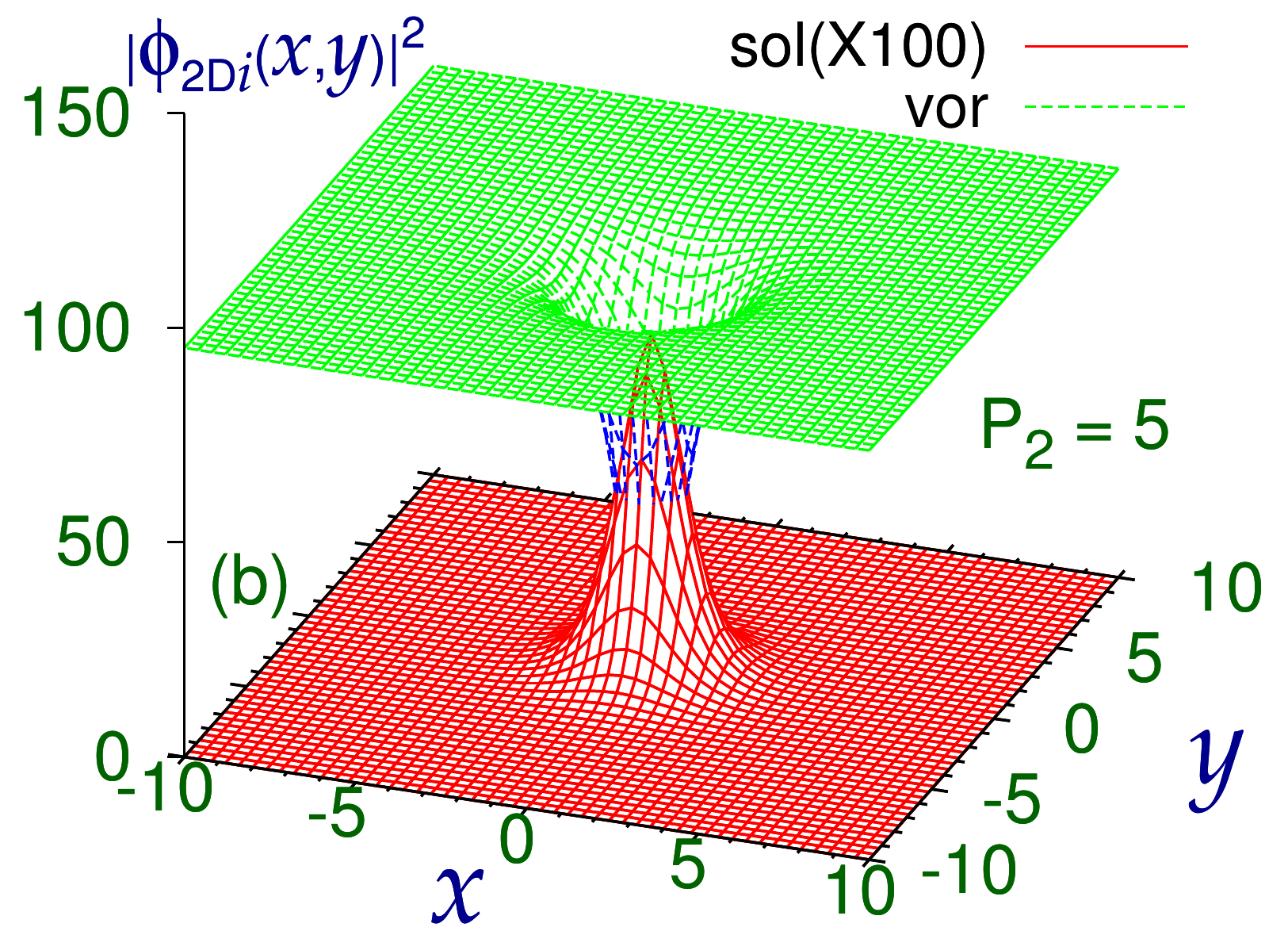}
\caption{(Color online)   Integrated 2D densities $|\phi_{{\mathrm2D}i}(x,y)|^2=\int_{t=-50}^{50} dt |\phi_i({\bf r})|^2$ for the soliton ($i=2$) and vortex ($i=1$) of the 3D spatiotemporal vortex-soliton with vortex power density 
 $p_1=1$ and  soliton powers (a) $P_2=10$  and (b) $P_2=5$ from a solution of Eqs. (\ref{eq5}) and (\ref{eq9}) in a cubic box of size $100^3$. The nonlinearity in the soliton  is self-focusing corresponding to the $+$ sign before $|\phi_2|^2$ in Eq. (\ref{eq5}).
For an easy visualization the densities of the soliton are multiplied by 50 and 100 in (a) and (b). 
}\label{fig4} 

\end{center}

\end{figure} 

In the  self-defocusing case, the 2D soliton has a larger spatial extension as can be seen from the density profiles shown   in Fig. \ref{fig3}(a) for $P_2= 1,10$ and 100. A large power $P_2$ corresponds to a 
large force on the first component hosting the vortex, which increases the size of the vortex core.  The densities in this case are qualitatively very similar to the  densities in a BEC vortex-soliton in the  self-defocusing case
as shown in the bottom row of Fig. 1 of Ref. \cite{law}.  
In this case there is no critical power $P_2$ for forming a stable soliton as in Fig. \ref{fig2}. { The vortex-soliton is stable for all self-defocusing  soliton powers $P_2$ and vortex power density  $p_1$. To demonstrate this numerically we plot in \ref{fig3}(b) the density profiles for 
soliton powers $P_2 =  10, 100,$ and 1000 for a small vortex power density $p_1=0.1$.  
Even a weak vortex with a small power density can support a soliton with a  very large 
self-defocusing power. However, the size of the vortex-soliton is much larger for a small 
vortex power density [compare the length scales in Figs. \ref{fig3}(a) and (b)].}
We have tested the stability of the these 
vortex-soliton beams numerically under a small perturbation (not presented here) in all cases in real-$z$ propagation. 
Next we will study the 3D  spatiotemporal case and will  discuss in details the question of stability numerically under a small perturbation.

 \begin{figure}[!t]

\begin{center}
\includegraphics[width=.8\linewidth]{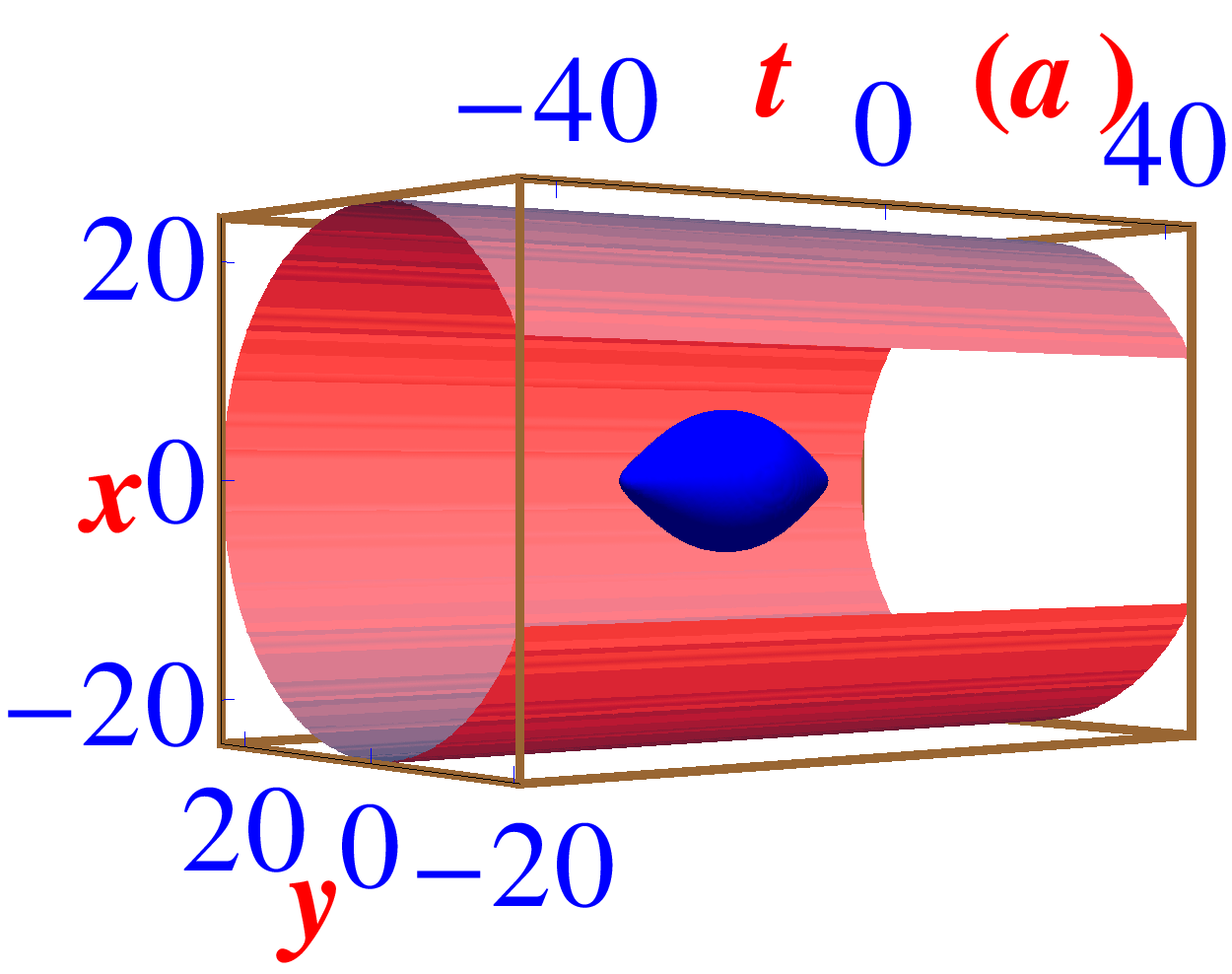}
 \includegraphics[width=.8\linewidth]{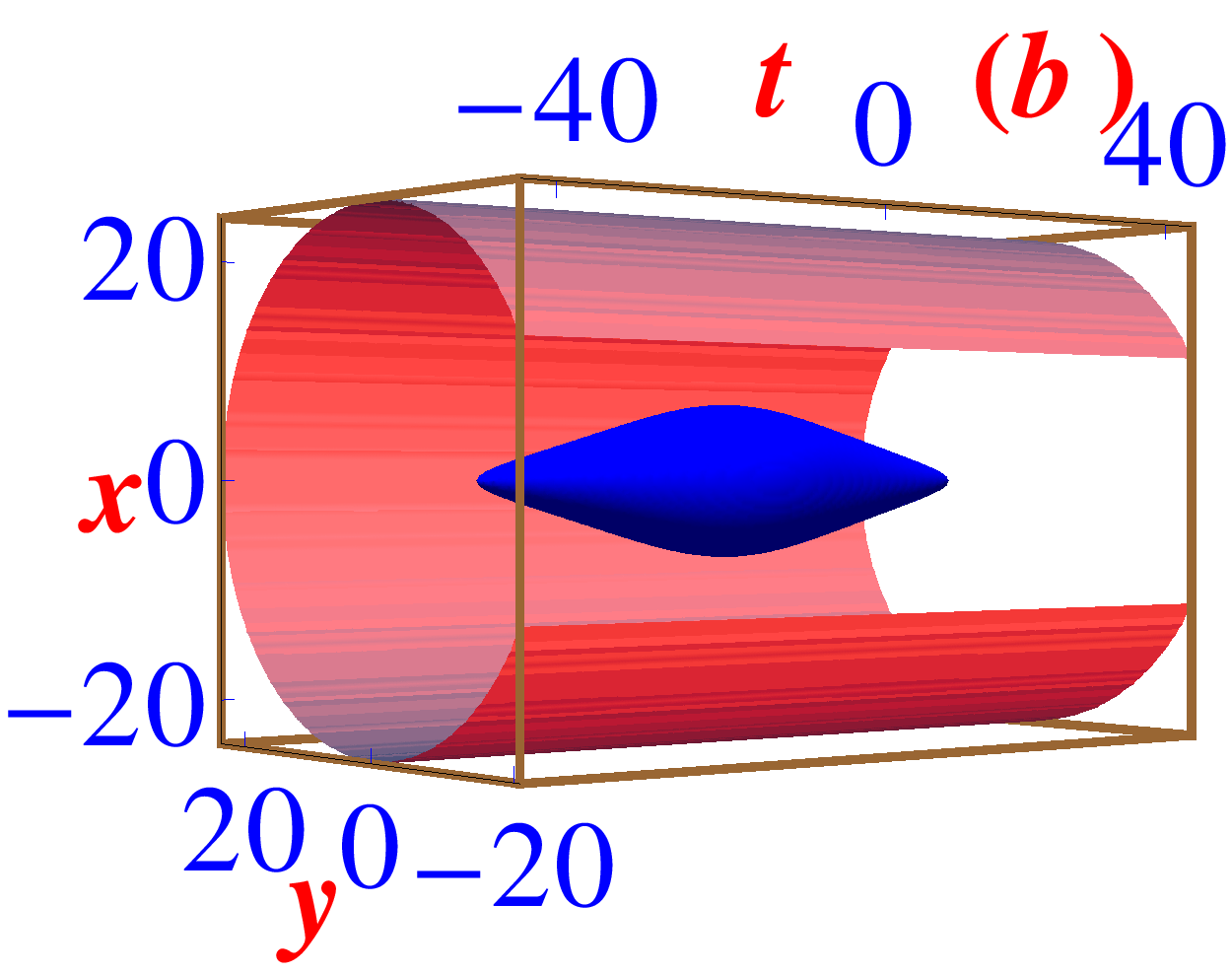} 

\caption{ (Color online)
3D isodensity contours $|\phi_i({\bf r})|^2/P_i$ of the binary vortex-soliton, showing
the vortex core (gray, pink in color) and the soliton (black, blue in color) profiles,
corresponding to (a) Fig.  \ref{fig4}(a) and  (b) Fig. \ref{fig4}(b).
Density on the contour is $10^{-6}$. The densities are divided by respective powers 
for the convenience of plotting in the same scale.
}\label{fig5} \end{center}

\end{figure}

In the 3D spatiotemporal case the  
numerical simulation is performed in the cubic box limited by  $x=y=t=\pm L, L= 50$ with 
powers $P_1=10^6$ and $P_2=10, 5$ for a self-focusing nonlinearity in the soliton, such that $\int_{-50}^{50} dx \int_{-50}^{50}dy  \int_{-50}^{50}dt |\phi_i(x,y,t)|^2 =P_i$.  The vortex power density in this case is $p_1\equiv P_1/(2L)^3 =1$.
A large power of the vortex beam is necessary for an efficient localization of the soliton by the repulsive centripetal force exerted by the vortex on the soliton in the $x-y$ plane. There is no such force in the temporal direction. We will see that the self-focusing nonlinearity of the soliton will be leading to a confinement of the soliton in time.   
The size of the vortex core is much smaller than the extension of the beam in the 
$x-y$ plane.

To visualize the  spatial localization of the   3D spatiotemporal vortex-soliton  in the $x-y$ plane we  calculated the integrated 2D density $|\phi_{{\mathrm2D}i}(x,y)|^2=\int dt |\phi_i({\bf r})|^2$ for the soliton and vortex. 
In Figs. \ref{fig4}(a) and (b) we plot the integrated 2D densities of the vortex-solitons for powers $P_2=10$, and $5$, respectively. The qualitative features of localization in the $x-y$ plane  is very similar to the localization of the spatial soliton in Fig. \ref{fig1}. 

Next we consider the full three-dimensional profile of the 3D spatiotemporal  vortex-soliton for 
illustrating its temporal localization.  In Figs. \ref{fig5}(a) and (b) we show the  isodensity contours $|\phi_i({\bf r})|^2/P_i$ of the binary vortex-soliton for powers $P_2=10$, and $5$, respectively. The length of the soliton in the temporal direction is larger in Fig. \ref{fig5}(b) compared to that in Fig. \ref{fig5}(a), due to a reduced self-attraction in the soliton for $P_2=5$ compared to the soliton with power  $P_2=10$, as
shown in Fig. \ref{fig5}(a). The temporal length of the soliton tends to infinity as the self-focusing power $P_2$ of the soliton reduces to zero, when the soliton cannot be localized in time. A self-focusing nonlinearity in the soliton is necessary for its localization.

 \begin{figure}[!t]

\begin{center}

 \includegraphics[width=\linewidth]{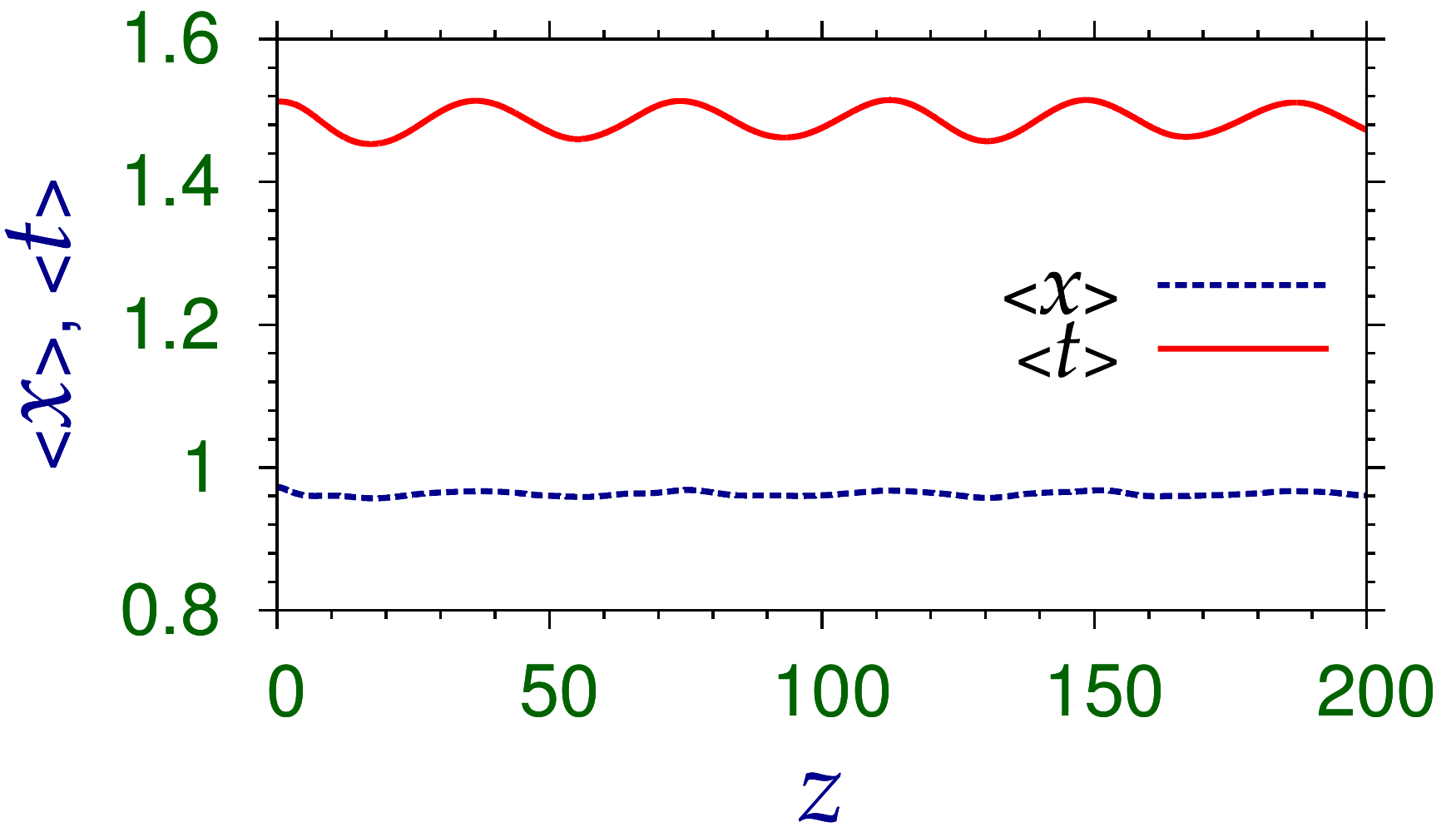}

\caption{ (Color online)
 RMS sizes $\langle x\rangle  ,\langle t \rangle$ during breathing oscillation of the
vortex-soliton of Fig. \ref{fig4}(a)    initiated by a sudden change of   soliton power from $P_2=10$ to 10.2 at $z=0$. 
}\label{fig6} \end{center}

\end{figure}

To demonstrate  the stability of the spatiotemporal vortex-soliton, we consider the one in Figs. \ref{fig4}(a) and \ref{fig5}(a) with soliton power $P_2=10$ 
and 
 subject the corresponding stationary state(s) obtained by  imaginary-$z$
propagation to real-$z$ propagation introducing a small perturbation, e.g., jumping 
the soliton power $P_2$ from $10$ to $10.2$ at $z=0$. Stable oscillation  of the resultant root-mean square  (RMS) sizes $\langle x\rangle =\langle y\rangle\ne \langle t\rangle$ of the soliton, 
illustrated  in Fig. \ref{fig6}, guarantees the stability of the binary vortex-soliton. The spatial and temporal sizes are different because of different dynamics  in space and time.

\begin{figure}[!t]

\begin{center}
\includegraphics[width=\linewidth]{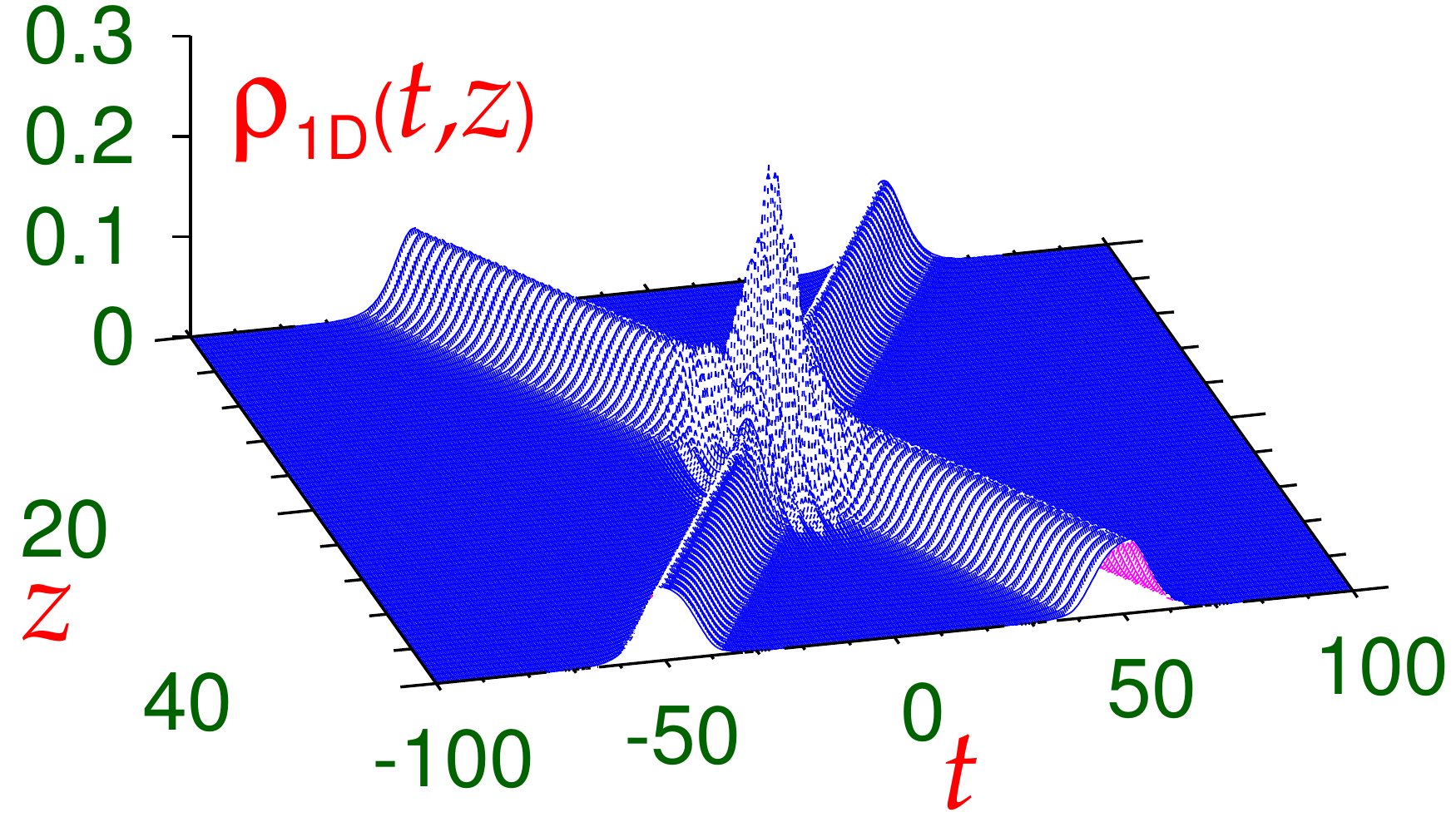}

\caption{ (Color online)
 The contour plot of  1D density   $\rho_{1D}(t,z)$   during  
 collision of two  spatiotemporal solitons of Fig. \ref{fig4}(b)  placed at $t=\pm 50$
at $z=0$. The solitons are set in motion with velocity $\sim 2.5$ in opposite directions so as to collide at $t=0$.      
}\label{fig7} \end{center}

\end{figure}

The collision between two integrable 1D solitons is truly elastic \cite{rmp,book}     and such solitons pass through each other without deformation.
The collision between  two 3D spatiotemporal solitons can at best be quasi-elastic. To test the solitonic nature of the present spatiotemporal solitons,
 we study the head-on collision of two solitons  
moving along the vortex core of the present vortex-soliton. The imaginary-$z$ profile of the binary vortex-soliton   of Fig. \ref{fig4}(b)  is used as the initial function in the real-$z$ simulation of collision, with two identical solitons  placed at $t=\pm 50$ initially for $z=0$.  To set the solitons in motion along the $t$ axis in opposite directions the soliton wave functions are multiplied by $\exp(\pm i v t), v=20$. To illustrate the dynamics upon real-$z$ simulation, we plot  the time evolution of  1D density $\rho_{1D}(t,z)\equiv \int dx\int dy |\phi(x,y,t,z)|^2$ in Fig. \ref{fig7}. 
The dimensionless velocity of a soliton   is $\sim 2.5$ and the deviation from elastic collision is found to be small.
  Considering the three-dimensional nature of collision, the distortion in the soliton profile is found to be negligible.

In 3D the vortex line of an isolated vortex is often found to suffer from transverse instability leading to a bending of vortex line \cite{pita}. No such instability is found in the present real-$z$ simulation. In the 3D spatiotemporal vortex-soliton the vortex beam applies a transverse centripetal force on the soliton in the $x-y$ plane thus stabilizing the soliton. The soliton also exerts a transverse centrifugal force on the vortex beam which prevents the vortex line from bending. 
The question of stability is the most critical in the spatiotemporal case as compared to the 2D spatial case.   
As the present spatiotemporal vortex-solitons are found to be stable, the same follows for the 2D spatial case, which could be  easier to realize experimentally.

\section{Summary and Discussion}

\label{IV}

Summarizing, we demonstrated the creation of a stable 2D spatial 
and a 3D spatiotemporal 
optical vortex-soliton in a binary system with the soliton moving in the core 
of the vortex. The nonlinearity in the vortex and the inter-beam nonlinearity are 
taken as self-defocusing Kerr type, whereas the  nonlinearity in the soliton in the 2D spatial case can be either self-focusing or self-defocusing Kerr type. However,  
the  nonlinearity in the soliton in the 3D spatiotemporal case 
can only be weakly self-focusing. 
  The soliton  is localized by a strong inter-beam 
repulsion. 
 This binary vortex-soliton  is a stable stationary state. 
In the 3D spatiotemporal case, the
optical soliton  can move with a constant velocity  along the vortex core in the temporal direction.   At medium velocities, the collision between the two 
spatiotemporal solitons moving along the vortex core 
 is   quasi elastic with 
no visible deformation of the final solitons.

An excellent account of different ways of generating a binary (or vector) soliton
experimentally is given in  chapter 9 of Ref. \cite{book}.  The experimental  observation  of a spatial binary (vector) optical soliton was accomplished nearly  two decades ago 
 \cite{1996}  in an AlGaAs slab waveguide, where the mutual trapping was achieved due to coupling between two polarization components of the beam.
The possibility of the observation of a binary vortex-soliton is considered in great detail in Ref. 
\cite{mus}. Their analysis is valid for both self-focusing and -defocusing nonlinearity in the soliton.     
The techniques of generating a 2D spatial optical vortex beam in a self-defocusing medium 
are  well known \cite{vorop,vorth}, hence   a  binary 
optical vortex-soliton could be within the current experimental possibilities.
{ The   binary NLS equations considered here also describe a 
binary BEC mixture \cite{law}, where the inter-species and intra-species interactions can be controlled independently by manipulating the different scattering lengths using  
optical \cite{opt} and magnetic \cite{mag} Feshbach resonances. In this fashion one can easily have defocusing inter-species nonlinearity and focusing nonlinearity in the soliton.
This will also provide a different testing ground for the analysis presented in this paper. }

\begin{acknowledgments}
Interesting discussion with Boris A. Malomed and F. K. Abdullaev is gratefully acknowledged. 
We thank the Funda\c c\~ao de Amparo 
\`a
Pesquisa do Estado de S\~ao Paulo (Brazil)  and  the
Conselho Nacional de Desenvolvimento   Cient\'ifico e Tecnol\'ogico (Brazil) for 
support.
\end{acknowledgments}

%

\begin{thebibliography}{99}

\bibitem{book}Y. S. Kivshar and G. Agrawal, {\it Optical Solitons: From Fibers to 
Photonic Crystals}, (Academic Press, San Diego, 2003).

\bibitem{sol}
Y. S. Kivshar and B. A. Malomed,
Rev. Mod. Phys. {\bf 61}, 763 (1989).

\bibitem{rmp}F. Dalfovo, S. Giorgini, S. Stringari, and L. P. Pitaevskii, 
Rev. Mod. Phys. {\bf 71}, 463 (1999);
F. K. Abdullaev, A. Gammal, A. M.  Kamchatnov, and L.  Tomio, 
Int. J. Mod. Phys. B {\bf 19}, 3415 (2005);
V. M. 
Perez-Garcia, H. Michinel, and H. Herrero,  Phys. Rev. A {\bf 57}, 3837
(1998).
 


\bibitem{dark}S. Burger, K. Bongs, S. Dettmer, W. Ertmer, K. Sengstock, A. Sanpera, G. V. Shlyapnikov, and M. Lewenstein, Phys. Rev. Lett. {\bf 83}, 5198 (1999);
 J. Denschlag {\it et al.}, Science {\bf 287}, 97 (2000).

 
 

\bibitem{temporal} P. Di Trapani, D. Caironi, G. Valiulis, A. Dubietis, R. Danielius, and  A. Piskarskas,  Phys. Rev. Lett. {\bf 81}, 570 (1998).
%
 
\bibitem{collapse}  Y. Silberberg,  Opt. Lett. {\bf 15}, 1282 (1990).

  
\bibitem{stempo}X. Liu, L. J. Qian, and F. W. Wise,  
Phys. Rev. Lett. {\bf 82}, 4631 (1999).

 
 \bibitem{arrest}D. E. Edmundson and R. H. Enns,  Opt. Lett. {\bf 17},
586 (1992); G. Fibich and B. Ilan,   Opt. Lett. {\bf 29},
887 (2004); L. Torner and Y. V. Kartashov, Opt. Lett.  {\bf 34}, 1129  
 (2009),
O. Bang, W. Krolikowski, J. Wyller, and J. J. Rasmussen,  Phys. Rev. E {\bf 66}, 046619 (2002).

\bibitem{modnon}D. E. Edmundson and R. H. Enns, Phys. Rev. A {\bf 51}, 2491 (1995); R. McLeod, K. Wagner, and S. Blair, Phys. Rev. A {\bf 52}, 3254
͑1995; A. A. Kanashov and A. M. Rubenchik, Physica D {\bf 4}, 122
͑1981͒; K. Hayata and M. Koshiba, Phys. Rev. E {\bf 51}, 1499 (1995).
  
 \bibitem{sadhan}S. K.  Adhikari, \pra
{\bf 69},  063613   (2004), \pre {\bf 70,}    036608 (2004);
H. Saito and M. Ueda,   Phys. Rev. Lett. {\bf 90}, 040403 (2003).

\bibitem{malomed1}
 F. K. Abdullaev, J. G. Caputo, R. A. Kraenkel, and B. A. Malomed,  Phys. Rev. A {\bf 67}, 013605 (2003).

 \bibitem{sadhan2}S. K.  Adhikari,
\pre {\bf 71,} 016611
 (2005).
 

\bibitem{malomed}
B. A. Malomed, D. Mihalache, F. Wise, and L. Torner,   J. Opt. B {\bf 7}, R53 (2005);
D. Mihalache, D. Mazilu, F. Lederer, B. A. Malomed, Y. V. Kartashov, L.-C.
Crasovan, and L. Torner,   Phys. Rev. E {\bf 73}, 025601(R) (2006);
A. Desyatnikov, A. Maimistov, and B.  Malomed,   Phys. Rev. E {\bf 61}, 3107
(2000);{
B. A. Malomed, F. Lederer, D. Mazilu, and D. Mihalache,   Phys. 
Lett. A {\bf 361}, 336 (2007).}

\bibitem{vorop}G. A. Swartzlander, Jr. and C. T. Law, \prl {\bf 69}, 2503 (1992);
L. Marrucci, C. Manzo, and D. Paparo, Phys. Rev. Lett. {\bf 96}, 163905 (2006);
M. D. Williams, M. M. Coles, D. S. Bradshaw, and D. L. Andrews; \pra {\bf 89}, 033837 (2014);
Y. Shen, G. T Campbell, B. Hage, H. Zou, B. C. Buchler, and P. K. Lam,
 J.   Opt. {\bf 15}, 044005 (2013).

\bibitem{vorth} A. S. Desyatnikov, L. Torner, and Y. S. Kivshar,
Prog. Opt. {\bf 47}, 291  (2005),  X.-T.  Gan, P. Zhang, S. Liu, F.-J. Xiao, 
and J.-L. Zhao, Chin. Phys. Lett. {\bf 25,} 3280 (2008);
K. Motzek, F. Kaiser,
J. R. Salgueiro, Y. Kivshar, and C. Denz,   Opt. Lett. {\bf 29}, 2285 (2004); C.
T. Law, X. Zhang, and G. A. Swartzlander,  Opt. Lett. {\bf 25}, 55 (2000).



  \bibitem{pita}L. P. Pitaevskii, Sov. Phys. JETP {\bf 13}, 451 (1961);
A. Aftalion and T. Riviere,
Phys. Rev. A {\bf 64}, 043611 (2001); 
A. Aftalion and I. Danaila,
Phys. Rev. A {\bf 68}, 023603 (2003).




\bibitem{law}K. J. H. Law, P. G. Kevrekidis, and L. S. Tuckerman, \prl {\bf 105}, 
160405 (2010).

\bibitem{mus}Z. H. Musslimani, M. Segev, D. N. Christodoulides, and M. Soljacic,
Phys. Rev. Lett. {\bf 84}, { 1164}  (2000).

\bibitem{pola}M. Pola, J. Stockhofe, P. Schmelcher, and P. G. Kevrekidis, \pra
{\bf 86}, 053601 (2012).



\bibitem{dark-bright}A. P. Sheppard and Y. S. Kivshar, Phys. Rev. E {\bf 55}, 4773  (1997).
 V. V.   Afanasjeb, E. M.  Dianov, and V. N.  Serkin, 
IEEE J. Quantum Electron  {\bf 25},  2656  (1989). 



\bibitem{CPC} P. Muruganandam and S. K. Adhikari, Comput. Phys.
Commun. {\bf 180}, 1888 (2009); J. Phys. B {\bf 36,} 2501 (2003);
D. Vudragovic, I. Vidanovic,
A. Balaz, P. Muruganandam, and S. K. Adhikari, Comput.
Phys. Commun. {\bf 183}, 2021 (2012). 



\bibitem{townes}R. Y. Chiao, E. Garmire, and C. H. Townes, Phys. Rev. Lett. {\bf 13}, 479 (1964).

\bibitem{berge}
L. Berg\'e, Phys. Rep. {\bf 303}, 259 (1998).


\bibitem{1996}J. U. Kang, G. I. Stegeman, and J. S. Aitchison, Opt. Lett. {\bf 21}, 189 (1996); J. U. Kang, J. S. Aitchison, G. I. Stegeman, and N. N. Akhmediev, Opt. Quantum Elec-
tron. {\bf 30}, 649 (1998).

\bibitem{opt}G. Thalhammer, M. Theis, K. Winkler, R. Grimm, and
J. Hecker Denschlag,   Phys. Rev. A {\bf 71}, 033403 (2005);
 M. Theis, G. Thalhammer, K. Winkler, M. Hellwig, G. Ruff,
R. Grimm, and J. Hecker Denschlag,   Phys. Rev. Lett. {\bf 93},
123001 (2004).

\bibitem{mag}S. Inouye, M. R. Andrews, J. Stenger, H.-J. Miesner, D. M.
Stamper-Kurn, and W. Ketterle,   Nature (London) {\bf 392}, 151 (1998).


\end{thebibliography}
\end{document}